\documentclass[journal,9pt]{IEEEtran}

\usepackage[utf8]{inputenc}
\usepackage[T1]{fontenc}

\usepackage{dsfont}
\usepackage{amsfonts,amsmath,mathrsfs}

\usepackage{apalike}
\usepackage{color}
\usepackage{psfrag}
\usepackage{graphicx} 
\usepackage{pstricks-add}

\usepackage{empheq}

\newtheorem{theorem}{Theorem}

\newtheorem{corollary}[theorem]{Corollary}

\newtheorem{definition}[theorem]{Definition}
\newtheorem{example}[theorem]{Example}

\newtheorem{lemma}[theorem]{Lemma}

\newtheorem{remark}[theorem]{Remark}

\DeclareMathOperator{\id}{\textrm{I}}
\DeclareMathOperator{\e}{\textrm{e}}
\DeclareMathOperator{\Zinv}{\mathcal Z_{\rm inv}}
\DeclareMathOperator{\real}{\mathbb R}

\newcommand{\eq}{\begin{equation}}
\newcommand{\qe}{\end{equation}}

\newcommand{\eqq}{$$\begin{aligned}}
\newcommand{\qqe}{\end{aligned}$$}

\newcommand{\bbm}{\begin{bmatrix}}
\newcommand{\ebm}{\end{bmatrix}}

\newcommand{\smallmat}[1]{\left[ \begin{smallmatrix}#1\end{smallmatrix} \right]}

\makeatletter
\renewcommand*\env@matrix[1][c]{\hskip -\arraycolsep
  \let\@ifnextchar\new@ifnextchar
  \array{*\c@MaxMatrixCols #1}}
\makeatother

\begin{document}

\title{
On the existence of strong functional observer 
}

\author{Micha\"el~Di~Loreto,
        Damien~\'Eb\'erard
\thanks{M. Di Loreto and D. {\'E}b\'erard are with INSA Lyon, Universit\'e Claude Bernard Lyon 1, Ecole Centrale de Lyon, CNRS, Amp\`ere,
UMR5005, 69621 Villeurbanne, France (e-mail: michael.di-loreto@insa-lyon.fr, damien.eberard@insa-lyon.fr)}
\thanks{Preprint submitted to IEEE Trans. on Automatic Control.}
}

\maketitle

\begin{abstract}
For arbitrary linear time-invariant systems, the existence of a strong functional observer is investigated. Such observer determines, from the available measurement on the plant, an estimate of a function of the state and the input. This estimate converges irrespective to initial state and input. This formulation encompass the cases of observer existence for known or unknown inputs and generalizes state-of-art. Necessary and sufficient conditions for such an existence are proposed, in the framework of state-space representation. These conditions are based on functional detectability property and its generalizations for arbitrary input, which include considerations on convergence of the estimation, irrespective to the initial state and the input. Known results on state detectability, input reconstruction or functional detectability are retrieved by particularizing the proposed conditions.
\end{abstract}


\section{Introduction}\label{section1}
This work addresses the observer issue as stated in \cite{hautus1983}. More precisely, LTI systems with two types of output are considered
\begin{subequations}\label{eq:LTI}
\begin{empheq}[left={\Sigma~:}\hspace{1em}]{align}
\dot x  =  Ax+Bu\,,	\label{eq:xdot} \\
y  =  Cx+Du\,,		\label{eq:y} \\
 z = Ex+Fu\,, 		\label{eq:z}
\end{empheq}
\end{subequations}
where $x\in\real^n$ is the state vector, $x(0)$ the initial state, $u\in\real^m$ is the input, $y\in\real^p$ is the output available for measurement, and $z\in\real^q$ is the output to be estimated. In \cite{hautus1983} the terminology \emph{$z$-observer based on $y$} is used. This is an observer which, using $y$ as input, gives an estimate $\hat z$ of $z$. Such an observer is required to be linear, causal and memoryless. It has a state-space realization in the form
\begin{subequations}\label{eq:z-obs}
\begin{empheq}[left={\Omega~:}\hspace{1em}]{align}
\dot{\xi} & =  G\xi+Hy\,, 	\label{eq:xidot}\\
\hat{z} & = Q\xi+Ry\,.	\label{eq:zhat}
\end{empheq}
\end{subequations}
Convergence of the error estimate $\hat z - z$ should be guaranteed for any initial state of both $\Sigma$ and $\Omega$, and for any input $u$. Precise statements on convergence will be defined later on.\\ 
In the present contribution, we investigate existence conditions of aforementioned observer, using structural properties of $\Sigma$ in the state-space framework. In~\cite{hautus1983}, a complete characterization of existence of (full) state observer has been proposed, irrespective to initial state and input, through the concept of strong detectability. In this paper, it has been further shown that existence of a \emph{$z$-observer based on $y$} is equivalent to the concept of strong functional detectability. The latter appears as a natural generalization of strong detectability to the case of a function of the state and the input to be estimated. However, such concept has not been characterized therein through structural and numerically reliable explicit conditions. Main purpose of the present contribution is to give such a characterization of strong functional detectability.\\
The chosen formulation, according to that introduced in~\cite{hautus1983}, allows us to encompass a wide variety of past contributions, including 
state functional observer with known or unknown input~\cite{luenberger1966,mooreledwich1975,schumacher1980,guansaif1991,houmuller1992,fernandotrinhjennings2010,trinhfernando2012,darouachfernando2020,darouachfernando2022}, simultaneous state-input observer~\cite{trinhfernando2012}, partial input reconstruction~\cite{bejarano2011}, inversion~\cite{moylan1977,xiongsaif2003,bejarano2009,diloretoeberard2022}, model matching~\cite{willems1982,malabrekucera1984,wonham1985} and disturbance decoupled estimation problems~\cite{basilemarro1992}. \\
State functional detectability for known input has been characterized in~\cite{moreno2005} through a frequency rank condition.  Still for known input, state functional detectability and design of a functional observer are studied in~\cite{darouach2000,darouachfernando2020} when the order of this observer corresponds to the dimension of the output to be estimated, and in~\cite{darouachfernando2022}. 
For unknown input, existence of full state observer is analyzed in~\cite{hautus1983}, while its design is performed in~\cite{houmuller1992,darouach2009,trinhfernando2012}.\\ 
In the case of unknown input,~\cite{darouach2004} investigates existence and design of a state functional observer with fixed order (corresponding to the dimension of the output to be estimated). The latter paper also assumes that the to be estimated output is input independent, as in~\cite{rotellazambettakis2016}. Existence characterization of a functional observer with unknown input in full generality has not been proposed, and this is precisely the main goal of the present contribution.\\ 
Starting from structural properties related to strong functional detectability and ensuing~\cite{hautus1983}, existence of a $z$-observer based on $y$ is characterized by necessary and sufficient conditions. Accordingly to the more recent aforementioned literature, terminology of strong functional observer is adopted instead of $z$-observer.  
We mainly develop two cases of strong functional observer, depending from estimation convergence property and its causal state-space realization. Particularizing the given existence conditions, we compare the obtained conditions with those of the literature, including the known input case, strong detectability, input reconstruction, and functional observer with unknown input and fixed order. \\
Main contributions include necessary and sufficient conditions for existence of a functional observer with known and unknown input in full generality, a unified problem formulation and analysis for (strong) functional detectability, and a
convergence insight of the estimation produced by such functional observer. 


\section{Notations}
\label{sec:notation}
\begin{subequations}
All signals are supposed to be impulsive-smooth distributions~\cite{hautussilverman1983,trentelmanetal2001}, so that their Laplace transform is well-defined. The normal rank of a rational matrix $M$ is defined as its maximal admissible rank
\eq
{\rm normrank}\,M = \max_{s\in\mathbb C} {\rm rank}\,M(s)\,.
\qe
Observe ${\rm rank}\,M(s) = {\rm normrank}\,M$ for all but finitely many $s$. Define the system matrices associated with \eqref{eq:LTI} as the two following polynomial matrices \cite{rosenbrock1970}
\eq
P(s)= \bbm s\id_n-A & -B \\ C & D \ebm \quad\text{and}\quad P_{\rm e}(s)= \bbm P(s)\\\begin{array}{cc} E & F \end{array} \ebm\,.
\qe
The invariant zeros of $P$ are the zeros of the nontrivial invariant polynomials in the diagonal Smith form of $P$. Thus an invariant zero is a value $s_0\in\mathbb C$ such that ${\rm rank}\,P(s_0) < {\rm normrank}\,P$. Denote by $\mathcal Z_{\rm inv}(P)$ the set of invariant zeros of $P$ (counting multiplicities). Let $\mathbb C^+$ be the set of complex numbers with nonnegative real part, and $\mathbb C^-$ its complementary set. Define $\mathcal Z_{\rm inv}^+(P) = \mathcal Z_{\rm inv}(P) \cap \mathbb C^+$. The output-decoupling zeros of $P$ are the $(A,C)$-unobservable eigenvalues of $A$. Thus an output-decoupling zero is a value $\lambda\in\mathbb C$ such that
\eq
{\rm rank}\bbm \lambda\id_n-A \\ C \ebm <n\,.
\qe
Denote by $\mathcal Z_{\rm od}(P)$ the set of output-decoupling zeros of $P$ (counting multiplicities). As previously, $\mathcal Z_{\rm od}^+(P)$ stands for the set $\mathcal Z_{\rm od}(P)\cap\mathbb C^+$.
Last recall a rational function in $s$ is said to be proper (resp. stable) if it admits a limit as $|s|\to\infty$ (resp. if its poles belong to $\mathbb C^-$).

\end{subequations}

\section{Preliminaries}

In~\cite{hautus1983}, the concepts of strong detectability and strong observability have been extended to deal with system $\Sigma$ given by~\eqref{eq:LTI}, having unknown input and an output $z$ to be estimated from output measurement $y$. Therein, the terminology used for output reconstruction was concerned with (stable integrating) $z$-observer based on $y$. In view of a more recent literature \emph{e.g.}~\cite{fernandotrinhjennings2010,trinhfernando2012,darouachfernando2022}, the terminologies \emph{functional detectability} and \emph{functional observer} have been introduced for the known input case (\textit{i.e.} $D=0$ in \eqref{eq:y} and $F=0$ in \eqref{eq:z}). Accordingly, the concepts introduced in \cite{hautus1983} are rephrased as follows.
\medskip
\begin{definition}\label{def:z-detectability} 
System $\Sigma$ given by \eqref{eq:LTI}  is said to be:
\begin{itemize}
\begin{subequations}
\item[(i)] strongly functional detectable if for any initial state $x(0)$ and any input $u$, one has
\eq\label{eq:sf-detectable}
y(t)=0\hspace{1em} (t>0) \Longrightarrow z(t)\to 0 \hspace{1em} (t\to\infty)\,.
\qe
\item[(ii)] strong$^\star$ functional detectable if for any initial state $x(0)$ and any input $u$, one has
\eq\label{eq:s*f-detectable}
y(t)\to 0 \hspace{1em} (t\to\infty) \Longrightarrow z(t)\to 0 \hspace{1em} (t\to\infty)\,.
\qe
\end{subequations}
\end{itemize}
\end{definition}
\begin{definition}\cite{hautus1983}
\label{def:obs}
System $\Omega$ given by \eqref{eq:z-obs} is said to be:
\begin{itemize}
\begin{subequations}
\item[(i)] a strong stable integrating functional observer for $\Sigma$ if there exist a nonnegative integer $k$ and a real $\tau>0$ such that, for any initial states of both $\Sigma$ and $\Omega$ and any input $u$, one has
\begin{equation}\label{eq:z-obs-2}
z(t)-\left(\tau \frac{\mathrm{d}}{\mathrm{d}t}+1\right)^k\hat{z}(t)\longrightarrow 0\quad \text{as}\quad t\longrightarrow \infty.
\end{equation}
\item[(ii)] a strong functional observer for $\Sigma$ if for any initial states of both $\Sigma$ and $\Omega$, and any input $u$, one has
\begin{equation}\label{eq:z-obs-1}
z(t)-\hat{z}(t)\longrightarrow 0\quad \text{as}\quad t\longrightarrow \infty.
\end{equation}
\end{subequations}
\end{itemize}
\end{definition}

\medskip
\begin{remark}\label{rem:f-detectable} 
Whenever the input $u$ is assumed to be known (thus including the zero input case), strongly and strong$^\star$ functional detectabilities are equivalent to \emph{functional detectability} as defined in the aforementioned literature, namely for any initial state $x(0)$ and $u=0$, property~\eqref{eq:sf-detectable} holds. Indeed, let $x_u$ be the forced state solution of~\eqref{eq:xdot}, depending only from the known input $u$. In the new coordinates $\tilde{x}=x-x_u$,~\eqref{eq:xdot} writes  
$\dot{\tilde{x}}=A\tilde{x}$, while~\eqref{eq:y} and~\eqref{eq:z} reduce to $\tilde y= y-Cx_u-Du = C\tilde{x}$ and $\tilde z = z-Ex_u-Fu = E\tilde{x}$, respectively. 
For such free input system, claimed equivalence is reached if \eqref{eq:sf-detectable} implies \eqref{eq:s*f-detectable} (the converse is trivially satisfied). For this, assume~\eqref{eq:sf-detectable} and the premise of~\eqref{eq:s*f-detectable}, \emph{i.e.} for any $\tilde{x}(0)$, $\tilde y(t) = C \mathrm{e}^{At} \tilde{x}(0)\rightarrow 0$ as $t\rightarrow\infty$. For any eigenvalue $\lambda$ of $A$, particularize $\tilde{x}(0)$ as its eigenvector to get $\tilde y(t) = \mathrm{e}^{\lambda t} C\tilde{x}(0)$. 
If $C\tilde{x}(0)=0$, $\tilde y = 0$ and conclusion follows from \eqref{eq:sf-detectable}. If $C\tilde{x}(0)\neq 0$, we get $\mathrm{Re}\,\lambda<0$, so that 
$\tilde z(t) = \mathrm{e}^{\lambda t} E\tilde{x}(0)$ converges to zero.\medskip
\end{remark}
It is worth noting that, in the general case, strong$^\star$ functional detectable implies strongly functional detectable, which in turn implies functional detectable. Both converses are false as illustrated by the following examples.

\begin{example}\label{ex:1} 
Consider $\Sigma$ given by the equations
\eq
\label{eq:ex:1}
\dot{x} = -x + \begin{bmatrix} 1 & -1 \end{bmatrix} u, \hspace{1em} y = x , \hspace{1em} z = x + \begin{bmatrix} 1 & 1 \end{bmatrix}u\,.
\qe
This system is functional detectable since for $u=0$ one has $y=z$, and thus \eqref{eq:sf-detectable} holds for any initial state. For arbitrary nonzero real $\alpha$, set  $u=\bbm \alpha & \alpha \ebm^T$ and take the zero initial state. Then $y=0$ but $z= 2\alpha \neq 0$. Thus \eqref{eq:sf-detectable} is not satisfied and the system is not strongly functional detectable.
\end{example}
\begin{example}\label{ex:2}
Consider $\Sigma$ given by the equations
\eq
\label{eq:ex:2}
\dot{x} = \begin{bmatrix} 0 & 0\\1&0 \end{bmatrix}x + \begin{bmatrix} 1 \\0 \end{bmatrix} u, \hspace{1em} y = \begin{bmatrix} 1 & 1 \end{bmatrix}x , \hspace{1em} z = u\,.
\qe
This system is strongly functional detectable since $y=0$ implies $u(t)= \e^{-t}u(0)$, and thus $z\to0$. 
However, the output $y(t) = t^{-1}\mathrm{sin}(t^2)$, which converges to zero, results from an input $u$ satisfying $\dot{u}+u=\ddot{y}$. Then the output $z=u$ has no asymptotical limit.  Thus \eqref{eq:s*f-detectable} is not satisfied and the system is not strong$^\star$ detectable.
\end{example}

\medskip
Last, recall the equivalence between strongly (or strong$^\star$) functional detectability, strong functional observer and the solvability of some algebraic matrix equation.
These results summarize as follows.\\

\begin{theorem}\cite[Theorem~3.8]{hautus1983} 
\label{thm:hautus-sf}
The following statements are equivalent:
\begin{enumerate}
\item $\Sigma$ given by \eqref{eq:LTI} is strongly functional detectable.
\item There exists a strong stable integrating functional observer $\Omega$ given by \eqref{eq:z-obs}.
\item There exist stable matrices $(M,N)$ such that
\eq
\label{eq:MN}
\bbm M(s) & N(s) \ebm P(s) = \bbm E & F \ebm\,.
\qe
\end{enumerate}
\end{theorem}
\begin{theorem}\cite[Theorem~3.2]{hautus1983}
\label{thm:hautus-s*f}
The following statements are equivalent:
\begin{enumerate}
\item $\Sigma$ given by \eqref{eq:LTI} is strong$^\star$ functional detectable.
\item There exists a strong functional observer $\Omega$ given by \eqref{eq:z-obs}.
\item There exist proper stable matrices $(M,N)$ such that \eqref{eq:MN} holds.
\end{enumerate}
\end{theorem}
\begin{remark}
\label{rem:z=Ny}
Observe equation \eqref{eq:MN} multiplied from the right by  $\smallmat{x(s) \\ u(s)}$ yields $z(s)= M(s) x(0) + N(s) y(s)$. Thus $\Omega$ defined by $\hat z(s) = N(s) y(s)$ with $N$ proper stable, has a causal realization in the form \eqref{eq:z-obs}. And further $z-\hat z = M(s)x(0)$ asymptotically vanishes since $M$ stable. System $\Omega$ is similarly designed when $N$ is stable by considering $(\tau s+ 1)^k \hat z(s)= N(s) y(s)$. See \cite{hautus1983} for details.
\end{remark}

\section{Strong functional detectability}

We now give a characterization of strong functional detectability by means of invariant zeros. Compared with the solvability of \eqref{eq:MN}, this characterization has the main advantages to be explicit, dependent only from parameters of $\Sigma$ in~\eqref{eq:LTI}, and to be numerically performed with efficient algorithm such as~\cite{vandoorendewilde1983}.  It is also worth noticing that, from Theorem~\ref{thm:hautus-sf}, such a characterization is equivalent to existence of a strong stable integrating functional observer $\Omega$ given by \eqref{eq:z-obs} for the system~$\Sigma$ in~\eqref{eq:LTI}.

\medskip

\begin{theorem} 		
\label{thm:sf-detectable}
System $\Sigma$ given by \eqref{eq:LTI} is strongly functional detectable if and only if
\begin{subequations}\label{eq:thm-sf}
\begin{empheq}[left={\left\{\hspace{1em}}, right={\right.}]{align}
&\mathrm{normrank}\,P = \mathrm{normrank}\,P_{\rm e} \label{eq:PPe}\\
&\mathcal{Z}_{\mathrm{inv}}^+(P)  = \mathcal{Z}_{\mathrm{inv}}^+(P_e).\label{eq:ZZe}
\end{empheq}
\end{subequations}
\end{theorem}
{\bf Proof.}
Applying Corollary~\ref{lem:gustafson} (in appendix) over the ring $\mathcal R$ of stable rational matrices, equation \eqref{eq:MN} has a solution in $\mathcal R$ if and only if 
$\smallmat{P \\ 0}$ and $P_{\e}$ are equivalent. Denote by $S$ the Smith form of $P$, that is
\eq
\begin{aligned}
S(s) & = Q(s) P(s) R(s) = \bbm \Lambda(s) & 0 \\ 0 & 0 \ebm\,,
\end{aligned}
\qe
where $Q,R$ are polynomial unimodular matrices (thus in $\mathcal R$) and $\Lambda$ is the diagonal matrix of invariant polynomials. The decomposition $\Lambda(s)= \Lambda^-(s)\Lambda^+(s)$, where $\Lambda^-,\Lambda^+$ are diagonal matrices such that the former only contains stable roots and has full rank in $\mathbb C^+$ and the latter only contains unstable roots, leads to the factorization
\eq
S(s) = S^-(s)S^+(s):= \bbm \Lambda^-(s) & 0 \\ 0 & \id \ebm \bbm \Lambda^+(s) & 0 \\ 0 & 0 \ebm \,,
\qe
with $S^-$ unimodular on $\mathcal R$. So $(S^-)^{-1}Q P R = S^+$ and therefore the matrices $\smallmat{P \\0}$  and  $\smallmat{S^+ \\ 0}$ are equivalent on $\mathcal R$. To conclude, it remains to see that \eqref{eq:thm-sf} holds if and only if the matrices $P_{\rm e}$ and $\smallmat{S^+ \\ 0}$ are equivalent on $\mathcal R$. The result follows by transitivity of matrix equivalence. \hfill $\Box$
\medskip

\begin{example}[example~\ref{ex:1} continued]
\label{ex:1bis}
Consider system $\Sigma$ defined by \eqref{eq:ex:1}. One can verify that $\Zinv(P)= \Zinv(P_{\rm e}) = \emptyset$, thus \eqref{eq:ZZe} holds. However since ${\rm normrank}\, P < {\rm normrank}\, P_{\rm e}$, equation \eqref{eq:PPe} is not fulfilled. Therefore the system is not strongly functional detectable.
\end{example}

\begin{example}[example~\ref{ex:2} continued]
\label{ex2bis}
Consider system $\Sigma$ defined by \eqref{eq:ex:2}. One can verify that $\Zinv(P)=\{-1\}$ and $\Zinv(P_{\rm e}) = \emptyset$, thus \eqref{eq:ZZe} holds. Moreover the normal ranks are equal. Therefore the system is strongly functional detectable.
\end{example}

\section{Strong$^\star$ functional detectability}

We now give a characterization of strong$^\star$ functional detectability. It relies on a characterization of the properness of solution of \eqref{eq:MN} by means of supremal invariant subspace.

Consider system $\Sigma$ given by \eqref{eq:LTI} and define its \emph{extended} system by the matrices
\begin{equation}\label{eq:AeBe}
A_e = \begin{bmatrix}A & B\\0&0\end{bmatrix},\hspace{1em} B_e = \begin{bmatrix}0\\\id_m\end{bmatrix},\hspace{1em} C_e = \begin{bmatrix}C&D\end{bmatrix}.
\end{equation}
Let $\mathscr{V}^\star_{C,D}$ stands for  the supremal $(A_e,B_e)$-invariant subspace contained in $ \mathrm{Ker}\,C_e$. It is well-known that it is the limit of the    algorithm, see \emph{e.g.} ~\cite{basilemarro1992,wonham1985}:
\begin{subequations}
\begin{align}
& \mathscr{V}^0_{C,D} =  \mathrm{Ker}\,C_e 		\label{eq:V0}\\
& \mathscr{V}^\mu_{C,D} =  \mathrm{Ker}\,C_e \cap A_e^{-1}({\rm Im}\,B_e+\mathscr{V}^{\mu-1}_{C,D})  \label{eq:Vmu}
\end{align}
\end{subequations}
where the set $A_e^{-1}\mathscr{V}$ is defined as $\left\{\psi\in\mathbb{R}^{n+m} : A_e\psi\in\mathscr{V}\right\}$. Remind this sequence is nonincreasing, and thus if $\mathscr{V}^{\mu+1}_{C,D}=\mathscr{V}^{\mu}_{C,D}$ then $\mathscr{V}^k_{C,D}=\mathscr{V}^{\mu}_{C,D}$ for all $k\geq\mu$. This algorithm converges in at most $\varrho = \mathrm{dim}(\mathrm{Ker}\,C_e)$ steps. Analogously, $\mathscr V^\star_{E,F}$ stands for the supremal $(A_{\rm e}, B_{\rm e})$-invariant subspace contained in ${\rm Ker}\,\bbm E & F \ebm$.
\medskip

\begin{theorem}\label{thm:s*f-detectable} 		
System $\Sigma$ given by \eqref{eq:LTI} is strong$^\star$ functional detectable if and only if \eqref{eq:thm-sf}  holds and in addition the set inclusion
\eq\label{eq:setinclusion}
\mathscr{V}^\star_{C,D}\cap {\rm Im}\,B_e \hspace{1em} \subseteq \hspace{1em} \mathscr{V}^\star_{E,F}\cap{\rm Im}\, B_e
\qe
is satisfied.
\end{theorem}
Once again, in virtue of Theorem~\ref{thm:hautus-s*f}, conditions in Theorem~\ref{thm:s*f-detectable} are equivalent to existence of a strong functional observer $\Omega$ for system~\eqref{eq:LTI}. Proof of Theorem~\ref{thm:s*f-detectable} requires a technical lemma (Lemma~\ref{lem:brandshautus} in appendix) and an intermediary result (Lemma~\ref{cor:brandshautus} below) on existence of a proper rational solution to~\eqref{eq:MN}.\\ 
For this, set $M_{C,D}^0= D$ and define for $k\geq 1$ the sequence of $(k+1)p\times(k+1)m$ lower block-triangular T\oe plitz matrices 
\eq
\label{eq:M_C,D}
M_{C,D}^k = 
\left[\ \vcenter{\hbox{$
  {\psset{linewidth=0.5pt, arrows=-,arrowinset=0.2, nodesep=3pt,colsep=0.3cm,rowsep=0.1cm,shortput=nab}
\begin{psmatrix}
D &   &   & & &  \\
CB & D  &   & & &  \\
CAB & CB  & D & & &  \\
  &   &   & & & \\ 
  &   &   & & &  \\
CA^{k-1}B &   & &  & CB & D 
\ncline{3,3}{6,6}\ncline{3,2}{6,5}
\ncline[linestyle=dashed]{3,1}{6,1}\ncline[linestyle=dashed]{6,1}{6,5}
\end{psmatrix}}
$}}\ \right]\,.
\qe
The matrix sequence $(M_{E,F}^k)_{k\geq 0}$ is defined in an analogous manner with $M_{E,F}^0=F$. Let $\mathbf{q}=(q_i)_{0\leq i\leq k}$, where $q_i\in\mathbb{R}^m$ and $k\in\mathbb{N}$. Then ${\bf q} \in \mathrm{Ker}\, M^k_{C,D}$ writes, for $i$ from $0$ to $k$ 
\eq
\label{eq:Mq}
\mathbf{e}_{i}^T M^k_{C,D} {\bf q} = \sum_{\ell=0}^{i-1}{CA^{i-1-\ell}Bq_\ell}+Dq_{i} = 0\,,
\qe
where ${\bf e}_i$ stands for the $i$-th $(p\times p)$ block-component of the canonical basis of $\mathbb{R}^{(k+1)p}$.
Observe \eqref{eq:Mq} can be rewritten as
\begin{subequations}
\eq
\label{eq:pi}
Cp_i + D q_i =0 \quad\text{with}\quad \left\{\begin{array}{l} p_0=0 \\ p_{i+1}= Ap_i + Bq_i \end{array} \right.
\qe
as well as 
\eq
\label{eq:psii}
C_e\psi_i = 0 \quad\text{with}\quad \left\{\begin{array}{l} \psi_0= B_e q_0 \\ \psi_{i+1}= A_e\psi_i + B_e  q_{i+1} \end{array} \right.
\qe
\end{subequations}
where $(A_e, B_e, C_e)$ are defined in \eqref{eq:AeBe}.  We shall now particularize Lemma~\ref{lem:brandshautus} for ${\bf A}= P$ and ${\bf B}= \bbm E & F \ebm $ over the ring $\mathcal R$ of proper rational matrices.

\medskip
\begin{lemma}\label{cor:brandshautus}
There exists ${\bf X}$ proper  such that ${\bf X} P = \bbm E & F \ebm$ if and only if one of the following equivalent conditions holds:
\begin{subequations}
\eq
\label{eq:brandshautusproper}
\forall p,q~: P(s) \smallmat{ p(s) \\ q(s)} \, \text{proper} \, \Longrightarrow  \bbm E & F\ebm \smallmat{ p(s) \\ q(s) } \, \text{proper}
\qe
\eq\label{eq:ker}
\forall k\in \mathbb N~: \hspace{1em} \mathrm{Ker}\,M_{C,D}^{k} \subseteq \mathrm{Ker}\,M_{E,F}^{k}
\qe
\eq\label{eq:V*}
\mathscr{V}^\star_{C,D}\cap {\rm Im}\,B_e \hspace{1em} \subseteq \hspace{1em} \mathscr{V}^\star_{E,F}\cap{\rm Im}\, B_e
\qe
\end{subequations}
\end{lemma}
{\bf Proof.}
\begin{subequations}
First, note there always exists a scalar polynomial ${\bf d}$ such that $P/{\bf d}$ is proper. Thus if \eqref{eq:brandshautus} holds with ${\bf A}= P/{\bf d}$ and ${\bf B}= \bbm E &F \ebm / {\bf d}$, then ${\bf X}(P/{\bf d}) = (\bbm E & F \ebm / {\bf d})$ is solved with ${\bf X}$ proper, and so does ${\bf X} P = \bbm E&F\ebm$ since $\mathcal R$ is an integral domain. Hence \eqref{eq:brandshautusproper}.\\
Second, observe $sp(s)-Ap(s)-Bq(s)$ proper implies $\deg q \geq 1+\deg p$. Thus the expansions\footnote{Higher order terms of $q$ are omitted since ${\rm Ker} \smallmat{-B \\ D} \subseteq {\rm Ker}\smallmat{ -B \\ F} $.}  (in decreasing powers) of $p$ and $q$ can be written as 
\begin{align}
\label{eq:p(s)q(s)}
p(s)&= \hspace{3em}  p_1 s^{k-1} + p_2 s^{k-2} + \dots \\
q(s)&=  q_0 s^{k}+  q_1 s^{k-1} + \dots 
\end{align}
with  $p_1, q_0\neq 0$. Then, by straightforward calculations,  $P \smallmat{p\\q} $ is proper if and only if \eqref{eq:pi} holds for $0\leq i \leq k$. That is $\smallmat{p\\q}\in\mathrm{Ker}\,M^k_{C,D}$.  Similarly $\bbm E & F \ebm\smallmat{p\\q}$ is proper, which reads $Ep_i + Fq_i =0$ for $0\leq i \leq k$, if and only if \eqref{eq:pi} holds { for $0\leq i \leq k$} replacing $(C,D)$ by $(E,F)$. That is $\smallmat{p\\q}\in  \mathrm{Ker}\, M^k_{E,F}$.  Hence \eqref{eq:ker} since $k$ is arbitrary.\\
Last, replace~\eqref{eq:pi} by \eqref{eq:psii}  in the above reasoning. Then $P \smallmat{p\\q} $ is proper if and only if $\psi_i \in \mathrm{Ker}\,C_e$ for $0\leq i \leq k$ (set $\psi_i= \smallmat{p_i \\ q_i}$). 
From a geometric viewpoint, this means that $\psi_0 \in \mathscr V_{C,D}^k$ defined in \eqref{eq:Vmu}. Since $k$ is arbitrary and $\psi_0=B_eq_0$, it follows that $\psi_0\in\mathscr V_{C,D}^\star \cap {\rm Im}  B_e$.
Similarly, $\bbm E & F \ebm \smallmat{p\\q}$ proper reads as \eqref{eq:psii} replacing $\mathrm{Ker}\, C_e$ by $\mathrm{Ker}\bbm E & F \ebm$. Hence \eqref{eq:V*}.  \hfill $\Box$
\end{subequations}
\medskip\\
{\bf Proof.} (of Theorem~\ref{thm:s*f-detectable})
It is obvious that if $\Sigma$ is strong$^\star$ functional detectable then \eqref{eq:MN} has a stable (resp. proper) solution hence \eqref{eq:thm-sf} by Theorem~\ref{thm:sf-detectable} (resp. \eqref{eq:setinclusion} by Lemma~\ref{cor:brandshautus}).\\
Conversely, we shall apply Lemma~\ref{lem:brandshautus} over the ring $\mathcal R$ of proper stable rational matrices. Note there always exists a scalar polynomial {\bf d} such that ${\bf A}=P/{\bf d} \in\mathcal R^{(n+p)\times(n+m)}$ and ${\bf B} = \bbm E & F \ebm/{\bf d}  \in\mathcal R^{r\times(n+p)}$.   Let $q\in \mathcal Q^{n+m}$. If ${\bf A}q \in\mathcal R^{n+p}$ then, in particular, ${\bf A}q$ is stable. This implies ${\bf B}q$ stable by \eqref{eq:thm-sf}. 
Indeed, \eqref{eq:thm-sf} characterizes strong functional detectability of $\Sigma$. Thus there exists $X$ stable such that $XP= \bbm E & F \ebm$.  Since the ring of stable rational matrices is an integral domain, it turns out that $X{\bf A}= {\bf B}$. Hence, by Lemma~\ref{lem:brandshautus}, ${\bf A}q$ stable implies ${\bf B}q$ stable.
Similarly,  if ${\bf A}q \in\mathcal R^{n+p}$ then ${\bf A}q$ is proper. This implies ${\bf B}q$ proper by \eqref{eq:setinclusion} (see Lemma~\ref{cor:brandshautus}).
In summary,  if  ${\bf A}q \in\mathcal R^{n+p}$ then  ${\bf B}q\in\mathcal R^r$. Thus \eqref{eq:MN} has a solution in $\mathcal R$ since $\mathcal R$ is an integral domain and the result follows.  \hfill $\Box$

\begin{remark}
It is worth noting that \eqref{eq:V*} is a finite set inclusion, whereas \eqref{eq:ker} is infinite. The former is thus numerically tractable and the latter not.
\end{remark}

\section{Discussion}

We shall now discuss particular choices of $\bbm E & F \ebm$, and show the present results encompass some literature ones.  

\subsection{State reconstruction: $[E \quad F]= [ {\rm I}_n \;\;\; 0 ] $ } \label{subsec:state}
\begin{subequations}
Consider $\Sigma$ given by \eqref{eq:LTI} with $\bbm E & F \ebm = \bbm \id_n & 0 \ebm$, \textit{i.e.} $z=x$. Then observe ${\rm rank}\, P_e = n+ {\rm rank} \smallmat{-B\\D}$ and thus $\mathcal Z_{\rm inv}(P_e) = \{\emptyset \}$. Therefore, according to Theorem~\ref{thm:sf-detectable},  $\Sigma$ is strongly functional detectable if and only if
\begin{align}
\mathrm{normrank}\,P & = n+ \mathrm{rank} \smallmat{-B \\ D} 		\label{eq:s-detectable1} \\
\mathcal Z_{\rm inv}(P) & \subseteq \mathbb C^- \,.				\label{eq:s-detectable2}
\end{align}
Those are precisely the conditions derived in \cite{hautus1983} for $\Sigma$ to be \emph{strongly detectable}. Moreover, direct computations show that~\eqref{eq:ker} (or equivalently~\eqref{eq:setinclusion}), that is $M^k_{C,D} {\bf q} =0 \Rightarrow M^k_{\id_n, 0}{\bf q} = 0$ for any $k\in\mathbb{N}$, simplifies as
\begin{align}\label{eq:ker-hautus}
\mathrm{Ker}\bbm D &0\\CB &D\ebm \subseteq \mathrm{Ker}\bbm 0 &0\\B &0\ebm.
\end{align}
Thus, according to Theorem~\ref{thm:s*f-detectable},  $\Sigma$ is strong$^\star$ functional detectable if and only if it is strongly functional detectable (\textit{i.e.} \eqref{eq:s-detectable1} and \eqref{eq:s-detectable2} hold) and in addition the kernel inclusion~\eqref{eq:ker-hautus} holds.
Those are precisely the conditions derived in \cite{hautus1983} for $\Sigma$ to be \emph{strong$^\star$ detectable}.
\end{subequations}

\subsection{Input reconstruction: $[E \quad F]= [ 0 \quad {\rm I}_m] $ }\label{subsec:input}

Consider $\Sigma$ given by \eqref{eq:LTI} with $\bbm E & F \ebm = \bbm 0 & \id_m \ebm$, \textit{i.e.} $z=u$. Then observe ${\rm rank}\, P_e = n+ m$ and thus $\mathcal Z_{\rm inv}(P_e) =\mathcal Z_{\rm od} (P)$ (see Section~\ref{sec:notation} for the notation).
According to Theorem~\ref{thm:sf-detectable}, $\Sigma$ is strongly functional detectable if and only if
\begin{subequations}
\begin{align}
\mathrm{normrank}\, P & = n+ m \label{eq:as-left1} \\
\mathcal Z_{\rm inv}(P)\backslash \{\mathcal Z_{\mathrm{od}}(P)&\cap \mathcal Z_{\rm inv}(P)\} \subseteq \mathbb C^-. \label{eq:as-left2}
\end{align}
Those are precisely the conditions derived in \cite{diloretoeberard2022} for $\Sigma$ to be \emph{asymptotically strongly left invertible}. Moreover, it is worth noting that $M^k_{0,\id_m} $ is the identity matrix. So the kernel inclusion $\mathrm{Ker}\, M^k_{C,D} \subseteq \mathrm{Ker}\, M^k_{0, \id_m}$ simply reads $\mathrm{Ker}\, M^k_{C,D} = \{ \emptyset \}$. Thus, according to Theorem~\ref{thm:s*f-detectable}, $\Sigma$ is strong$^\star$ functional detectable is and only if it is strongly functional detectable (\textit{i.e.} \eqref{eq:as-left1} and \eqref{eq:as-left2} hold) and in addition 
\eq
{\rm rank}\, D = m\,.
\qe 
\end{subequations}
We retrieve the conditions derived in \cite[Theorem 15]{diloretoeberard2022} for $\Sigma$ to be  asymptotically \emph{strong$^\star$ left invertible}.

\subsection{State functional reconstruction with known input}\label{subsec:statefunctional}
Consider system $\Sigma$ given by \eqref{eq:LTI} with $\bbm E & F \ebm = \bbm E & 0 \ebm$.
Under the assumptions that $\Sigma$ is either controllable or with unstable uncontrollable modes, and the input $u$ is known, it is claimed in~\cite[Theorem~4]{darouachfernando2022} that there exists a functional observer if and only if the system is functional detectable (see Remark~\ref{rem:f-detectable}), that is
\eq
\label{eq:fd-Darouach}
\mathrm{rank}\begin{bmatrix}s\id_n-A\\C\end{bmatrix}=
\mathrm{rank}\begin{bmatrix}s\id_n-A\\C\\E\end{bmatrix},\; \forall \;s\in\mathbb{C}^+.
\qe
This result has the main advantage to provide an asymptotic functional observer design, but it has the main drawback to only apply to a specific subclass of LTI system (with known input). For instance, 
\eq
\label{eq:ex}
\dot x = \bbm -1&0 \\ 0 & -1 \ebm x\,,\quad y =x\,,\quad z=\bbm 1 & 0 \ebm x\,.
\qe
is not controllable but is internally stable. Thus it does not fit within assumptions in \cite[Theorem~4]{darouachfernando2022}. However, system $\Omega$ given by \eqref{eq:z-obs} with null matrices for $G,H,Q$ and $R=\bbm 1&0 \ebm$ defines a (strong) functional observer for \eqref{eq:ex}. In contrast, existence of such an observer can be studied with the results presented in the current paper. Indeed, straightforward computations show that ${\rm normrank}\,(P) = {\rm normrank}\,(P_e)$ and $\mathcal Z_{\rm inv} (P)= \mathcal Z_{\rm inv}(P_e)$. Thus \eqref{eq:PPe} and \eqref{eq:ZZe} are satisfied and the system is strongly functional detectable by Theorem~\ref{thm:sf-detectable}. In addition, since $\mathscr V^\star_{C,D} \cap {\rm Im} B_e$ is the null subspace, the set inclusion \eqref{eq:setinclusion} is trivially satisfied. Hence $\Sigma$ is strong$^\star$ functional detectable, and there exists a strong functional observer for \eqref{eq:ex} by Theorem~\ref{thm:s*f-detectable}. \\
Actually, in the case of known input (or zero input, see Remark~\ref{rem:f-detectable}), there exists a strong functional observer if and only if~\eqref{eq:thm-sf} are satisfied (see Theorem~\ref{thm:sf-detectable}).  In such a case, setting $B$ and $D$ as zeros matrices,  it is seen that~\eqref{eq:thm-sf} implies~\eqref{eq:fd-Darouach}, but the converse is false.

\subsection{State functional reconstruction with fixed order}\label{subsec:statefunctionalfixedorder}
\begin{subequations}\label{eq:Darouach2004}
Consider system $\Sigma$ given by \eqref{eq:LTI} with $\bbm E & F \ebm = \bbm E & 0 \ebm$. A variety of past contributions dealt with a functional observer of fixed order corresponding to $\mathrm{dim}(z)=q$. These contributions include \emph{e.g.}~\cite{darouach2000} for known input and~\cite{darouach2004} for unknown input. In the latter paper, the case $\bbm E & F \ebm = \bbm E & 0 \ebm$ is analyzed. But as the direct feedthrough term $F$ takes little effort to be considered into such analysis, \cite[Theorem 2]{darouach2004} is rephrased as follows. For a controllable system $\Sigma$, there exists a strong$^\star$ functional observer with order $\mathrm{dim}(z)=q$ if and only if 
\eq
\label{eq:Darouach2004proper}
{\small\mathrm{Ker}\begin{bmatrix}E & F &0 \\C & D & 0\\CA &CB&D\end{bmatrix}\subseteq \mathrm{Ker}\begin{bmatrix}EA&EB&F\end{bmatrix}}
\qe
and for all $s\in\mathbb{C}^+$
\eq
\label{eq:Darouach2004stable}
{\small\mathrm{rank}\begin{bmatrix}E(s\id_n-A) &-EB&0\\C&D&0\\CA&CB&D\end{bmatrix} = \mathrm{rank}\begin{bmatrix}E & F &0 \\C & D & 0\\CA &CB&D\end{bmatrix}}.
\qe
A direct verification shows that~\eqref{eq:Darouach2004proper} implies~\eqref{eq:ker}, and~\eqref{eq:Darouach2004stable} implies $\mathrm{rank}\,P(s)=\mathrm{rank}\,P_e(s)$ for all $s\in\mathbb{C}^+$, that is~\eqref{eq:thm-sf} is fulfilled. Nevertheless, both converses are false since~\eqref{eq:Darouach2004} is concerned with a fixed order functional observer. For instance, the plant   
$$
\dot{x}=\begin{bmatrix}0&1\\0&0\end{bmatrix}x+\begin{bmatrix}1\\1\end{bmatrix}u,\;
y=u,\;z=\begin{bmatrix}1&-1\end{bmatrix}x
$$
does not verify~\eqref{eq:Darouach2004proper}, while~\eqref{eq:ker} holds. Actually, this plant is not functional detectable nor strongly functional detectable, but~\eqref{eq:MN} admits a proper (but not stable) solution. Similarly, the plant  
$$
\dot{x}=\begin{bmatrix}0&1&0\\0&0&1\\0&0&1\end{bmatrix}x,\;
y=\begin{bmatrix}1&0&0\end{bmatrix}x,\;z=\begin{bmatrix}0&1&0\end{bmatrix}x
$$
does not satisfy~\eqref{eq:Darouach2004stable}, while it is strong$^\star$ functional detectable (Theorem~\ref{thm:s*f-detectable} holds). 
\end{subequations}

\section{Conclusion}
Existence of a strong functional observer, namely for any initial state and any input, has been fully characterized for LTI systems. Given necessary and sufficient existence conditions are provided within a state-space approach. These conditions are numerically tractable for computations, and generalize classical frequency detectability conditions.
Future works may be devoted to the design of this strong functional observer, and in particular to its minimal order.

\section{Appendix}

Let $\mathcal{R}$ be a principal-ideal domain with quotient field $\mathcal{Q}$, such that $\mathcal{R}\neq \mathcal{Q}$.
\begin{lemma} \cite[Lemma 3.7]{brandshautus1982}
\label{lem:brandshautus}
Let $\mathbf{A}\in \mathcal{R}^{n\times m}$ and $\mathbf{B}\in \mathcal{R}^{r\times m}$. There exists $\mathbf{X}\in \mathcal{R}^{r\times n}$ such that $\mathbf{XA}=\mathbf{B}$ if and only if 
\eq\label{eq:brandshautus}
\forall q \in \mathcal Q^m~: \hspace{1em}
\mathbf{A}q\in \mathcal{R}^n \;\; \Longrightarrow \;\; \mathbf{B}q\in \mathcal{R}^r.
\qe
\end{lemma}

\begin{corollary}
\label{lem:gustafson}
\cite[Theorem 1]{gustafson1979}
Let $\mathbf{A}\in \mathcal{R}^{n\times m}$ and $\mathbf{B}\in \mathcal{R}^{r\times m}$. There exists $\mathbf{X}\in \mathcal{R}^{r\times n}$ such that $\mathbf{XA} = \mathbf{B}$ if and only if
there exist unimodular matrices $\mathbf{U}\in\mathcal{R}^{(n+r)\times(n+r)}$ and $\mathbf{V}\in\mathcal{R}^{m\times m}$  such that
\begin{equation}
\label{eq:gustafson}
\mathbf{U}\begin{bmatrix}\mathbf{A}\\0\end{bmatrix}\mathbf{V} = \begin{bmatrix}\mathbf{A}\\\mathbf{B}\end{bmatrix}.
\end{equation}
\end{corollary}

\bibliography{autosam}

\begin{thebibliography}{}

\bibitem[Basile and Marro, 1992]{basilemarro1992}
Basile, G. and Marro, G. (1992).
\newblock {\em Controlled and conditioned invariants in linear system theory}.
\newblock Prentice-Hall, Englewood Cliffs, NJ.

\bibitem[Bejarano, 2011]{bejarano2011}
Bejarano, F.~J. (2011).
\newblock Partial unknown input reconstruction for linear systems.
\newblock {\em Automatica}, 47:1751--1756.

\bibitem[Bejarano et~al., 2009]{bejarano2009}
Bejarano, F.~J., Fridman, L., and Poznyak, A. (2009).
\newblock Unknown input and state estimation for unobservable systems.
\newblock {\em SIAM J. on Control and Optimization}, 48:1155--1178.

\bibitem[Brands and Hautus, 1982]{brandshautus1982}
Brands, J.~J.~A.~M. and Hautus, M.~L.~J. (1982).
\newblock Asymptotic properties of matrix differential operators.
\newblock {\em J. Math. Analysis and Appl.}, 87:199--218.

\bibitem[Darouach, 2000]{darouach2000}
Darouach, M. (2000).
\newblock Existence and design of functional observers for linear systems.
\newblock {\em IEEE Trans. on Automatic Control}, 45:940--943.

\bibitem[Darouach, 2004]{darouach2004}
Darouach, M. (2004).
\newblock Functional observers for systems with unknown inputs.
\newblock In {\em 16th Int. Symposium on Mathematical Theory of Networks and
  Systems}, Leuven, Belgium.

\bibitem[Darouach, 2009]{darouach2009}
Darouach, M. (2009).
\newblock Complements to full order observer design for linear systems with
  unknown inputs.
\newblock {\em Applied Mathematics Letters}, 22:1107--1111.

\bibitem[Darouach and Fernando, 2020]{darouachfernando2020}
Darouach, M. and Fernando, T. (2020).
\newblock On the existence and design of functional observers.
\newblock {\em IEEE Trans. on Automatic Control}, 65:2751--2759.

\bibitem[Darouach and Fernando, 2023]{darouachfernando2022}
Darouach, M. and Fernando, T. (2023).
\newblock Functional detectability and asymptotic functional observer design.
\newblock {\em IEEE Trans. on Automatic Control}, 68:975--990.

\bibitem[{Di~Loreto} and Eberard, 2023]{diloretoeberard2022}
{Di~Loreto}, M. and Eberard, D. (2023).
\newblock Strong left inversion of linear systems and input reconstruction.
\newblock {\em IEEE Trans. on Automatic Control}, 68:3612--3617.

\bibitem[Fernando et~al., 2010]{fernandotrinhjennings2010}
Fernando, T.~L., Trinh, H.~M., and Jennings, L. (2010).
\newblock Functional observability and the design of minimum order linear
  functional observers.
\newblock {\em IEEE Trans. on Automatic Control}, 55:1268--1273.

\bibitem[Guan and Saif, 1991]{guansaif1991}
Guan, Y. and Saif, M. (1991).
\newblock A novel approach to the design of unknown input observers.
\newblock {\em IEEE Trans. on Automatic Control}, 36:632--635.

\bibitem[Gustafson, 1979]{gustafson1979}
Gustafson, W.~H. (1979).
\newblock Roth's theorems over commutative rings.
\newblock {\em Linear Algebra and its Applications}, 23:245--251.

\bibitem[Hautus, 1983]{hautus1983}
Hautus, M.~L.~J. (1983).
\newblock Strong detectability and observers.
\newblock {\em Linear Algebra and its Applications}, 50:353--368.

\bibitem[Hautus and Silverman, 1983]{hautussilverman1983}
Hautus, M.~L.~J. and Silverman, L.~M. (1983).
\newblock System structure and singular control.
\newblock {\em Linear Algebra and its Applications}, 50:369--402.

\bibitem[Hou and Muller, 1992]{houmuller1992}
Hou, M. and Muller, P.~C. (1992).
\newblock Design of observers for linear systems with unknown inputs.
\newblock {\em IEEE Trans. on Automatic Control}, 37:871--875.

\bibitem[Luenberger, 1966]{luenberger1966}
Luenberger, D.~G. (1966).
\newblock Observers for multivariable systems.
\newblock {\em IEEE Trans. on Automatic Control}, 11:190--197.

\bibitem[Malabre and Ku\v{c}era, 1984]{malabrekucera1984}
Malabre, M. and Ku\v{c}era, V. (1984).
\newblock Infinite structure and exact model matching problem : A geometric
  approach.
\newblock {\em IEEE Trans. on Automatic Control}, 29:266--268.

\bibitem[Moore and Ledwich, 1975]{mooreledwich1975}
Moore, J.~B. and Ledwich, G.~F. (1975).
\newblock Minimal order observers for estimating linear functions of a state
  vector.
\newblock {\em IEEE Trans. on Automatic Control}, 20:623--632.

\bibitem[Moreno, 2005]{moreno2005}
Moreno, J.~A. (2005).
\newblock Simultaneous observation of linear systems: A state-space approach.
\newblock {\em IEEE Trans. on Automatic Control}, 50:1021--1025.

\bibitem[Moylan, 1977]{moylan1977}
Moylan, P.~J. (1977).
\newblock Stable inversion of linear systems.
\newblock {\em IEEE Trans. on Automatic Control}, 22:74--78.

\bibitem[Rosenbrock, 1970]{rosenbrock1970}
Rosenbrock, H.~H. (1970).
\newblock {\em State space and multivariable theory}.
\newblock Wiley, New-York.

\bibitem[Rotella and Zambettakis, 2016]{rotellazambettakis2016}
Rotella, F. and Zambettakis, I. (2016).
\newblock A direct design procedure for linear state functional observers.
\newblock {\em Automatica}, 70:211--216.

\bibitem[Schumacher, 1980]{schumacher1980}
Schumacher, J.~M. (1980).
\newblock On the minimal stable observer problem.
\newblock {\em Int. J. Control}, 32:17--30.

\bibitem[Trentelman et~al., 2001]{trentelmanetal2001}
Trentelman, H., Stoorvogel, A.~A., and Hautus, M.~L.~J. (2001).
\newblock {\em Control theory for linear systems}.
\newblock Springer-Verlag, in Communications and Control Engineering, London.

\bibitem[Trinh and Fernando, 2012]{trinhfernando2012}
Trinh, H. and Fernando, T. (2012).
\newblock {\em Functional observers for dynamical systems}.
\newblock Springer, in Lecture Notes in Control and Information Sciences n.420,
  Berlin.

\bibitem[{Van Dooren} and Dewilde, 1983]{vandoorendewilde1983}
{Van Dooren}, P. and Dewilde, P. (1983).
\newblock The eigenstructure of an arbitrary polynomial matrix: {C}omputational
  aspects.
\newblock {\em Linear Algebra and its Applications}, 50:545--579.

\bibitem[Willems, 1982]{willems1982}
Willems, J.~C. (1982).
\newblock Almost invariant subspaces : An approach to high gain feedback
  design. {Part II} : Almost conditionally invariant subspaces.
\newblock {\em IEEE Trans. on Automatic Control}, 27:1071--1085.

\bibitem[Wonham, 1985]{wonham1985}
Wonham, W.~M. (1985).
\newblock {\em Linear multivariable control : A geometric approach}.
\newblock Springer-Verlag (3rd Ed.), New-York.

\bibitem[Xiong and Saif, 2003]{xiongsaif2003}
Xiong, Y. and Saif, M. (2003).
\newblock Unknown disturbance inputs estimation based on a state functional
  observer design.
\newblock {\em Automatica}, 39:1389--1398.

\end{thebibliography}
\bibliographystyle{apalike}        

\end{document}